\documentclass[a4paper,11pt]{article}
\pdfoutput=1 

\usepackage{jinstpub} 
\usepackage{graphicx}
\usepackage{color}
\usepackage{amsmath}
\usepackage{verbatim}
\usepackage{changepage}
\usepackage{graphicx}
\usepackage{epstopdf}
\AppendGraphicsExtensions{.gif}

\usepackage{lineno}
\usepackage{mdwlist}
\usepackage{subfigure}
\usepackage{xspace}
\usepackage{marginnote}
\usepackage{array}
\usepackage[utf8]{inputenc}
\usepackage{bibentry}
\usepackage{appendix}
\usepackage{acronym}
\usepackage{hepnicenames}
\usepackage{xcolor,bytefield}
\usepackage[colorinlistoftodos,prependcaption,textsize=tiny]{todonotes}

\hypersetup{
 pdftitle={},%
 pdfauthor={},%
 pdfsubject={},%
 pdfkeywords={},%
 pdfstartview={},%
 bookmarksopen=true, breaklinks=true, debug=true,
 colorlinks=true, linkcolor=blue, citecolor=blue, urlcolor=blue}

\title{The DAQ system of the 12,000 Channel CMS High Granularity Calorimeter Prototype}
\author[3]{B.~Acar,}
\author[2]{G.~Adamov,}
\author[32]{C.~Adloff,}
\author[23]{S.~Afanasiev ,}
\author[40]{N.~Akchurin,}
\author[3,5]{B.~Akg\"{u}n,}
\author[22]{M.~Alhusseini,}
\author[6]{J.~Alison,}
\author[4]{G.~Altopp,}
\author[9]{M.~Alyari,}
% ...
\affiliation[2]{Georgian Technical University, \\ 77 Kostava Str 0175, Tbilisi, Georgia}
\affiliation[3]{Bogazici University, \\Bebek 34342, Istanbul, Turkey}
\affiliation[4]{Brown University, \\182 Hope Street, Providence 02912, RI, USA}
\affiliation[5]{CERN,\\Espl. des Particules 1, 1211 Geneve 23 Switzerland}
\affiliation[6]{Carnegie Mellon University, \\ 5000 Forbes Ave, Pittsburgh 15213, PA, USA}
% ...
\affiliation[9]{Fermilab,\\ Wilson Road, Batavia 60510, IL, USA}
% ...
\affiliation[22]{The University of Iowa,\\ 203 Van Allen Hall, Iowa City, 52242, Iowa, USA}
\affiliation[23]{International Intergovernmental Organization Joint Institute for Nuclear Research JINR, \\ 6 Joliot-Curie St, Dubna 141980, Moscow, Russia}
% ...
\affiliation[32]{National Central University Taipei (NCU),\\No.300, Jhongda Rd 32001, Jhongli City Taiwan}
% ...
\affiliation[40]{Texas Tech University,\\ Lubbock 79409, TX, USA}
% ...

\emailAdd{Roger.Rusack@cern.ch, Bora.Akgun@cern.ch}

\newcommand{\colorbitbox}[3]{%
    \rlap{\bitbox{#2}{\color{#1}\rule{\width}{\height}}}%
    \bitbox{#2}{#3}}
\definecolor{lightblue}{rgb}{0.59,1,1}
\definecolor{lightcyan}{rgb}{0.84,1,1}
\definecolor{lightyellow}{rgb}{0.94,1,0.23}
\definecolor{lighteryellow}{rgb}{0.97,1,0.6}
\definecolor{lightmagenta}{rgb}{1,0.39,1}
\definecolor{lightermagenta}{rgb}{1,0.74,1}
\definecolor{lightorange}{rgb}{1,0.76,0.28}
\newcommand{\fakethirtytwobits}[1]{%
  \tiny
  \ifnum#1=1234567890
  #1
  \else
  \ifnum#1>10
  \count32=#1
  \advance\count32 by 16
  \the\count32%
  \else
  \ifnum#1<5
  #1%
  \else
  \ifnum#1=6
  $\cdots$%
  \else
  \ifnum#1=9
  $\cdots$%
  \fi
  \fi
  \fi
  \fi
  \fi
}

\abstract{The CMS experiment at the CERN LHC will be upgraded to accommodate the 5-fold increase in the instantaneous luminosity expected at the High-Luminosity LHC (HL-LHC)~\cite{bib.PhaseII_TP}. Concomitant with this increase will be an increase in the number of interactions in each bunch crossing and a significant increase in the total ionising dose and fluence. One part of this upgrade is the replacement of the current endcap calorimeters with a high granularity sampling calorimeter equipped with silicon sensors, designed to manage the high collision rates~\cite{bib.HGCAL_TDR}. As part of the development of this calorimeter, a series of beam tests have been conducted with different sampling configurations using prototype segmented silicon detectors. In the most recent of these tests, conducted in late 2018 at the CERN SPS, the performance of a prototype calorimeter equipped with ${\approx}12,000\rm{~channels}$ of silicon sensors was studied with beams of high-energy electrons, pions and muons.  This paper describes the custom-built scalable data acquisition system that was built with readily available FPGA mezzanines and low-cost Raspberry PI computers.}


\begin{document}
\maketitle
\flushbottom
\section{Introduction}
\label{sec:intro}

\begin{figure}
  \centering
  \includegraphics[width=.9\linewidth]{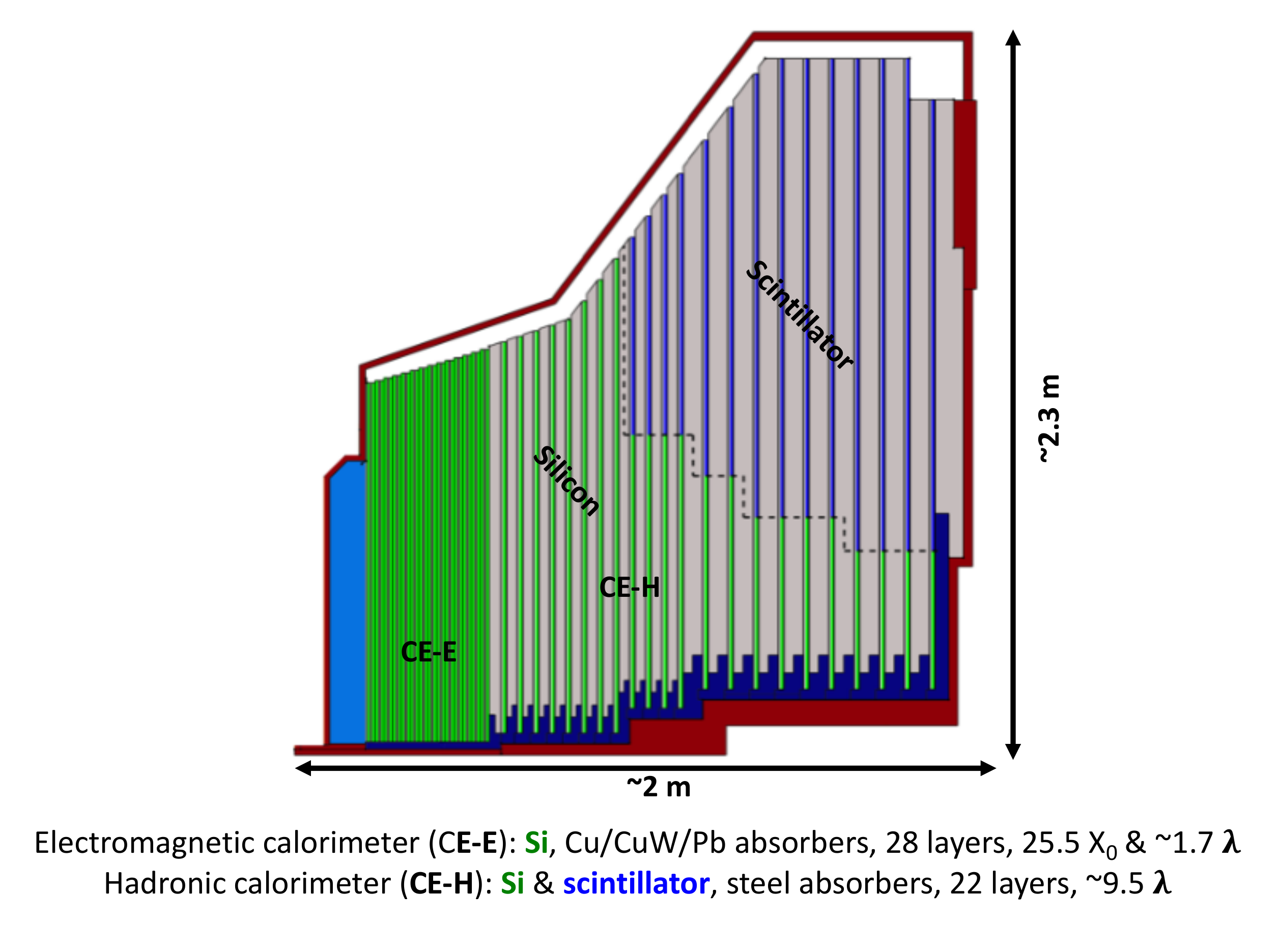}
  \caption{\label{fig:cms-hgcal} Schematic view of the CMS high granularity Endcap Calorimeter.}
\end{figure}

The HL-LHC at CERN is planned to operate with an instantaneous luminosity of $5 \times 10^{34}$~cm$^{-2}$s$^{-1}$ or higher, delivering up to ten times more integrated luminosity than is expected in the current LHC programme. This increase poses significant challenges in the design and operation of the detectors at the HL-LHC. In particular, in the forward direction the absorbed dose will be up as much as 2 MGy and the fluence in the innermost region is expected to reach $10^{16}$~n$_{eq}$/cm$^2$, which is an unprecedented level in high energy collider experiments. Additionally there will be ${\approx} 140$ proton-proton interactions occurring (pile up) in every bunch crossing, which happens at a rate of 40\,MHz. This considerably complicates the reconstruction of events. To contend with these conditions, the CMS Collaboration is planning a series of upgrades to some of the existing detector components, and replacing others with new detectors designed specifically to mitigate the effect of the high pile up~\cite{bib.PhaseII_TP}.  As part of this upgrade programme the current electromagnetic and hadronic calorimeters in the endcaps will be replaced with a new calorimeter, known as the `High-Granularity Calorimeter' (HGCAL)~\cite{bib.HGCAL_TDR}.  This new sampling calorimeter (CE) will be sub-divided into two sections, the electromagnetic (CE-E) and the hadronic (CE-H), in an arrangement shown in Figure~\ref{fig:cms-hgcal}. The CE-E will be equipped with silicon sensors, while the CE-H will be equipped with both silicon sensors and scintillator tiles read out directly with SiPMs. In the CE-H the silicon sensors will be at small radii, close to the beam where the radiation levels are highest, and scintillator tiles at the larger radii.  The absorber of CE-E will be a mixture of lead, copper and sintered copper-tungsten, while in CE-H the absorber plates will be stainless steel. The hexagonal silicon sensors will be subdivided into hexagonal cells with areas of ${\approx} 1.1\,\rm{cm}^2$ or 0.5~cm$^2$, with the sensors with smaller cells placed at small radii. The full calorimeter will be operated at -30$^{\circ}$C to reduce the dark current in the silicon sensors and in the SiPMs. There will be 28 sampling layers in CE-E and 22 in CE-H. This high-transverse granularity, combined with the high longitudinal segmentation of the calorimeter has been selected, within the constraints of cost and available space, to optimise the identification and measurement of hadronic and electromagnetic showers in the presence of the high pile up.

The basic detector unit is a silicon module. A module consists of a silicon sensor, glued to a baseplate on one side and to a printed circuit board (PCB) for the readout on the opposite side. The individual cells of the sensor are connected electrically to a readout ASIC on the PCB with wire-bonds that pass through holes in the PCB. The ASIC amplifies and digitises the analogue signals and transmits them to the off-detector electronics on receipt of an external trigger signal.  By 2018 more than 100 prototype silicon modules have been produced using 6-inch hexagonal silicon sensors subdivided into cells with areas of ${\approx}$1.1~cm$^{2}$. A module is shown in Fig. \ref{fig:module_interposer} (left). Further details of the construction and assembly of the silicon modules used in these tests may be found in~\cite{bib.H2}. The modules were assembled into a prototype of HGCAL that was tested with beams of electrons, pions and muons at the CERN SPS. 

The ASIC used to read out of the signals from the silicon cells was the Skiroc2-CMS~\cite{bib.SKIROC2_CMS}, a custom ASIC designed by the OMEGA group at Ecole Polytechnique. The function of this 64-channel ASIC was to measure both the amplitude of the signal and its time of arrival, so not only the energy response can be measured, but the timing performance can be characterised. This Skiroc2-CMS ASIC has many of the features of the HGCAL front-end readout ASIC in development. 

In the latest beam test in late 2018 the HGCAL prototype was equipped with 94-hexagonal silicon modules arranged into a 26 radiation length electromagnetic section and 5 nuclear interaction length hadronic section. Behind the prototype calorimeter we placed the Analog Hadronic Calorimeter (AHCAL) prototype, developed by the CALICE Collaboration~\cite{bib.ahcal}. This calorimeter is a scintillator-based sampling calorimeter, similar in design to the proposed design of the HGCAL~\cite{bib.HGCAL_TDR}, but with much finer longitudinal segmentation. In the final test at the CERN SPS, data were taken with beams of muons, charged hadrons and electrons with energies ranging from 20 to 300~GeV at the H2 beam line of the CERN-SPS over a period of two weeks in October 2018. 

The data acquisition (DAQ) system for the beam tests needed to be flexible and scalable to control and read out the increasing number of prototype silicon modules as they became available. It was designed with readily available FPGA mezzanines and low-cost Raspberry PIs, and scaled up to work with ${\approx}$12,000 channels for the final test.    

This paper describes the DAQ system and is structured as follows: the overall architecture of the system is described in section~\ref{sec:overview}; the data format and the back-end DAQ components are described in section~\ref{sec:backend}; the DAQ software is explained in section~\ref{sec:software_framework}; in section~\ref{sec:synchronization} the detector systems used for system synchronisation is discussed and the operational experience is discussed in section~\ref{sec:performance}.
\section{DAQ System Architecture}
\label{sec:overview}

Each hexagonal cell of the silicon sensor was connected to the 64-channel Skiroc2-CMS ASIC. Each channel of this ASIC had a low noise pre-amplifier followed by high- and low-gain shapers, with a shaping time of 40~ns, and time-over-threshold (ToT) and time-of-arrival (ToA) circuits. Both  of the shapers had analogue-to-digital converters (ADCs) that sampled the signal every 25~ns.  It also had a circuit to measure the ToA of large amplitude signals (${}>3 \rm{fC}$) with a precision of 50~ps. 
Further details of the design can be found in~\cite{bib.SKIROC2_CMS}.

\begin{figure}[htbp]
\centering 
\includegraphics[width=.613\textwidth,origin=c,angle=0]{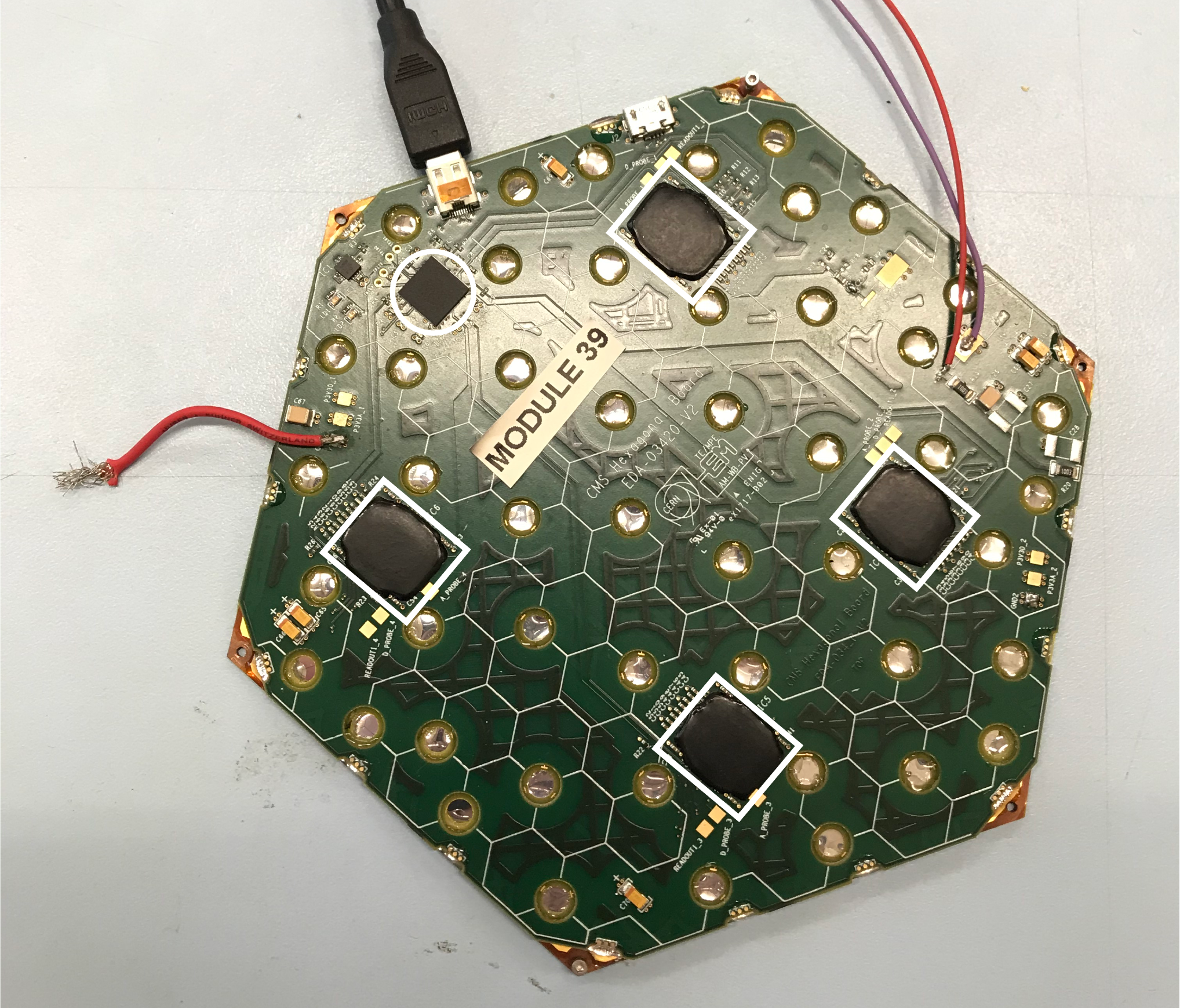}
\includegraphics[width=.38\textwidth,origin=c,angle=0]{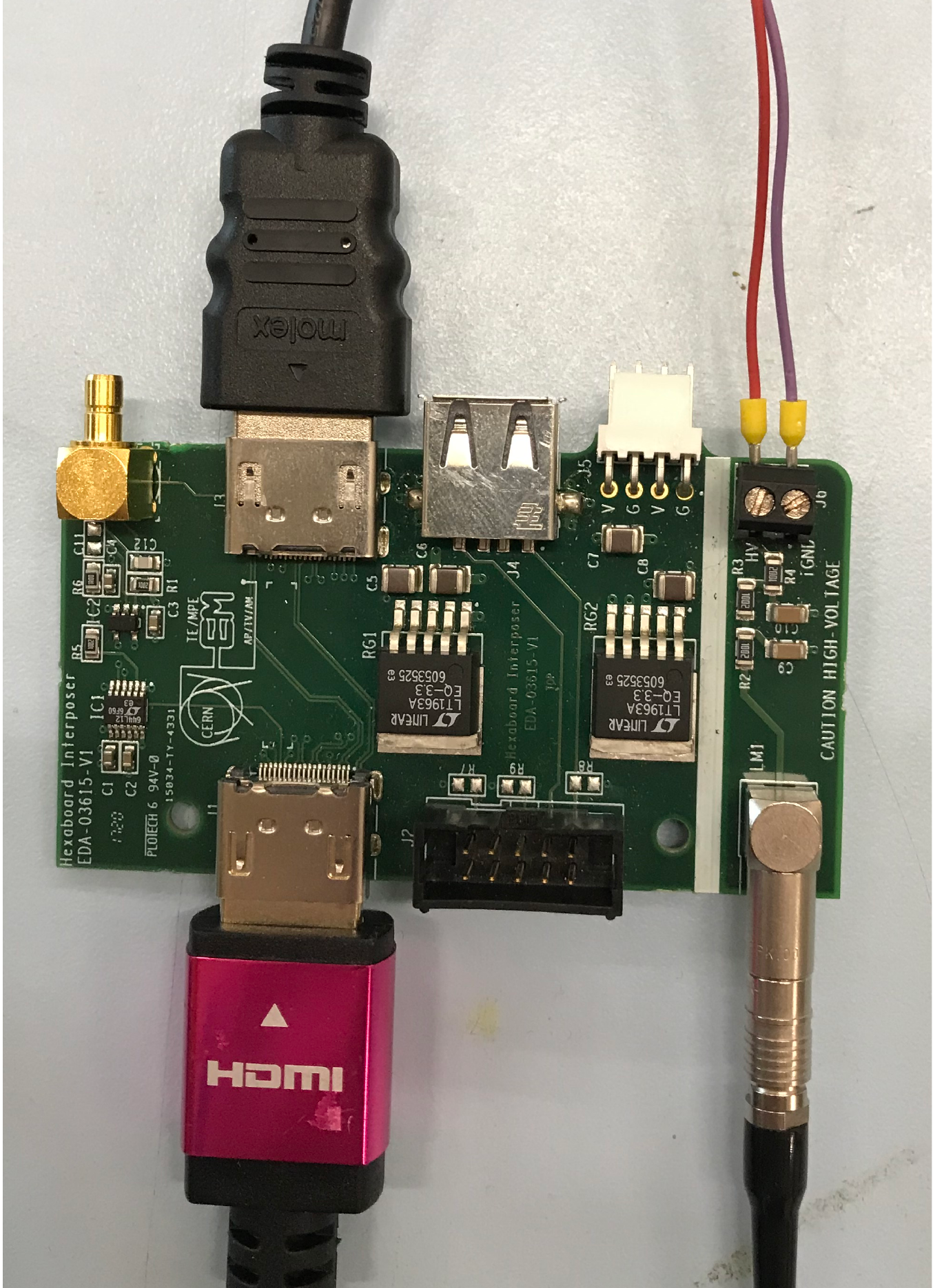}
\caption{\label{fig:module_interposer} (left) A prototype module used in beam tests. The Skiroc2-CMS ASICs are marked with white rectangles. The MAX\textregistered10 FPGA is marked with a white circle. The red grounding wire (on the left side) and, red and violet bias wires (on the right side) are soldered on the PCB. The micro HDMI (uHDMI) cable is connected on the top. (right) An interposer board used in beam tests. The HDMI cable and bias voltage wires, connecting the interposer board to the prototype module, are connected on the top. The HDMI and RG174~cables, connecting the interposer board to the DAQ board, are connected on the bottom.}
\end{figure}

To simplify routing of the signals in a very dense board, four Skiroc2-CMS ASICs were used to readout the 128 channels of each silicon sensor, leaving half of the channels unused. The PCB also had a MAX\textregistered10 field programmable gate array (FPGA) to control the readout of the module. It received the clock, trigger and busy signals from the off-detector electronics, aggregated the data from the Skiroc2-CMS ASICs and transmitted it to the off-detector electronics. Figure~\ref{fig:module_interposer}~(left) is a photograph of a prototype module, with the Skiroc2-CMS ASICs marked with  rectangles, and the MAX\textregistered10 FPGA, on the top left, is indicated with a white circle.  

Each of the prototype silicon modules was connected to the off-detector DAQ boards through an interposer board. These boards regulated the 5~V output from the DAQ boards to the 3.3~V needed by the prototype modules via HDMI cables. They also filtered and transmitted the bias voltage coming from the DAQ boards through RG174~cables to wires soldered to the prototype modules. A photograph of an interposer board is shown in Figure~\ref{fig:module_interposer}~(right). 

\begin{figure}[htbp]
\centering 
\includegraphics[width=.95\textwidth,origin=c,angle=0]{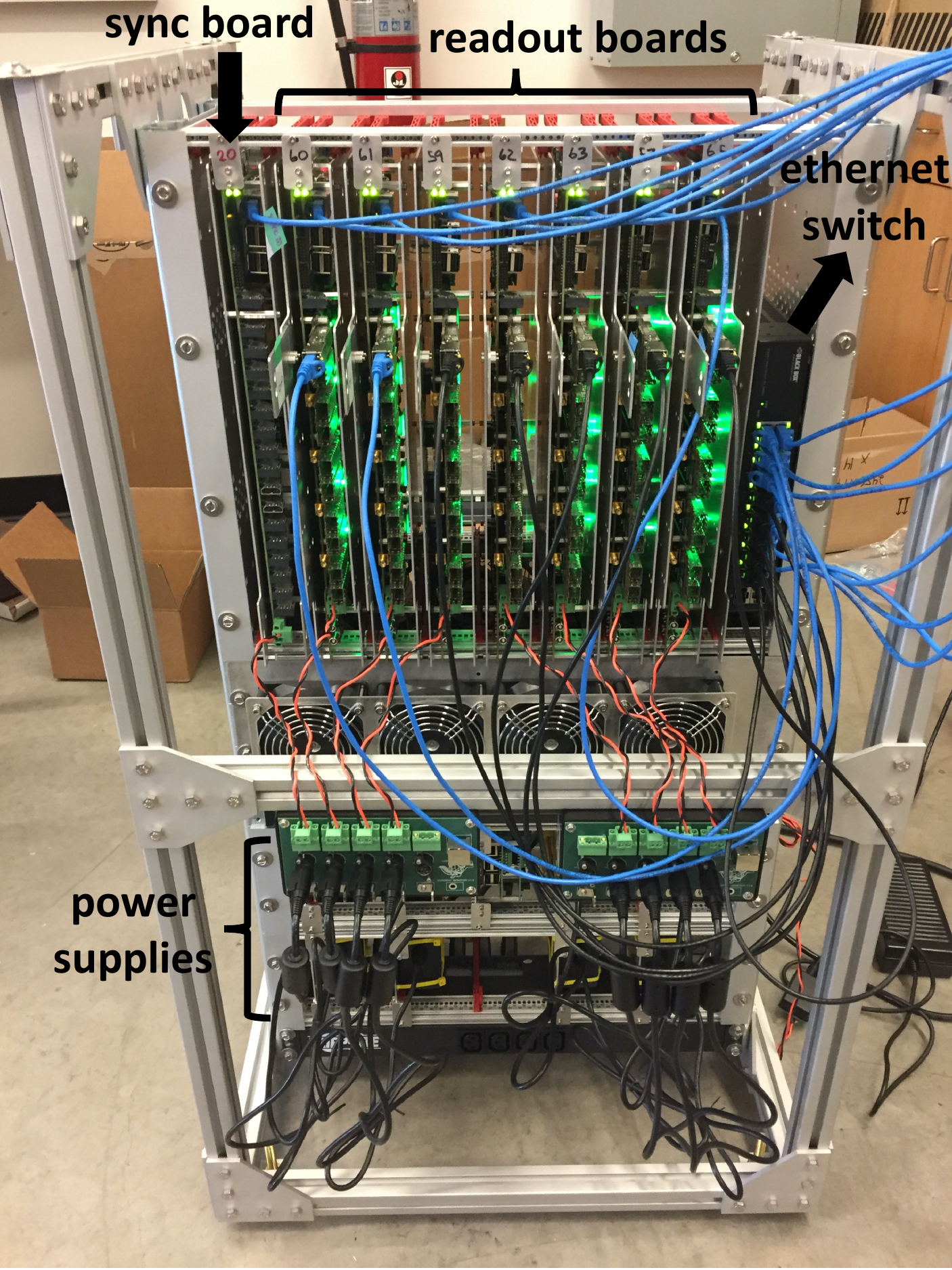}
\caption{\label{fig:crate} A crate, used in HGCAL beam tests, populated with one sync board and seven readout boards. 
The sync board (in the leftmost slot) distributed the clock, trigger and busy signals to the readout boards via HDMI cables, not shown. The readout boards were used to send control data to the silicon modules, to receive data from them, and to transmit the data to the acquisition computer through an Ethernet switch (on the right side of the readout boards). 
The readout and sync boards were powered by the power modules located under the DAQ crate.}
\end{figure}

The off-detector electronics consisted of a set of custom 9U readout boards, each of which could receive data from up to eight silicon modules. All the the readout boards were controlled by a single custom 9U synchronisation board, the `sync board'. The sync board distributed the clock, trigger and busy signals to all the readout boards. The readout boards communicated with the data acquisition computer through Ethernet, with a 100 Mbit/s output of the readout boards connected to a Gigabit Ethernet switch, from which data were sent to the DAQ computer for processing. In the tests with 94 silicon modules, one sync board and fourteen readout boards were mounted in two custom air-cooled crates. The crate that was equipped with one sync board and seven readout boards is shown in Figure~\ref{fig:crate}. 

\begin{figure}[htbp]
\centering 
\includegraphics[width=1.05\textwidth,origin=c,angle=0]{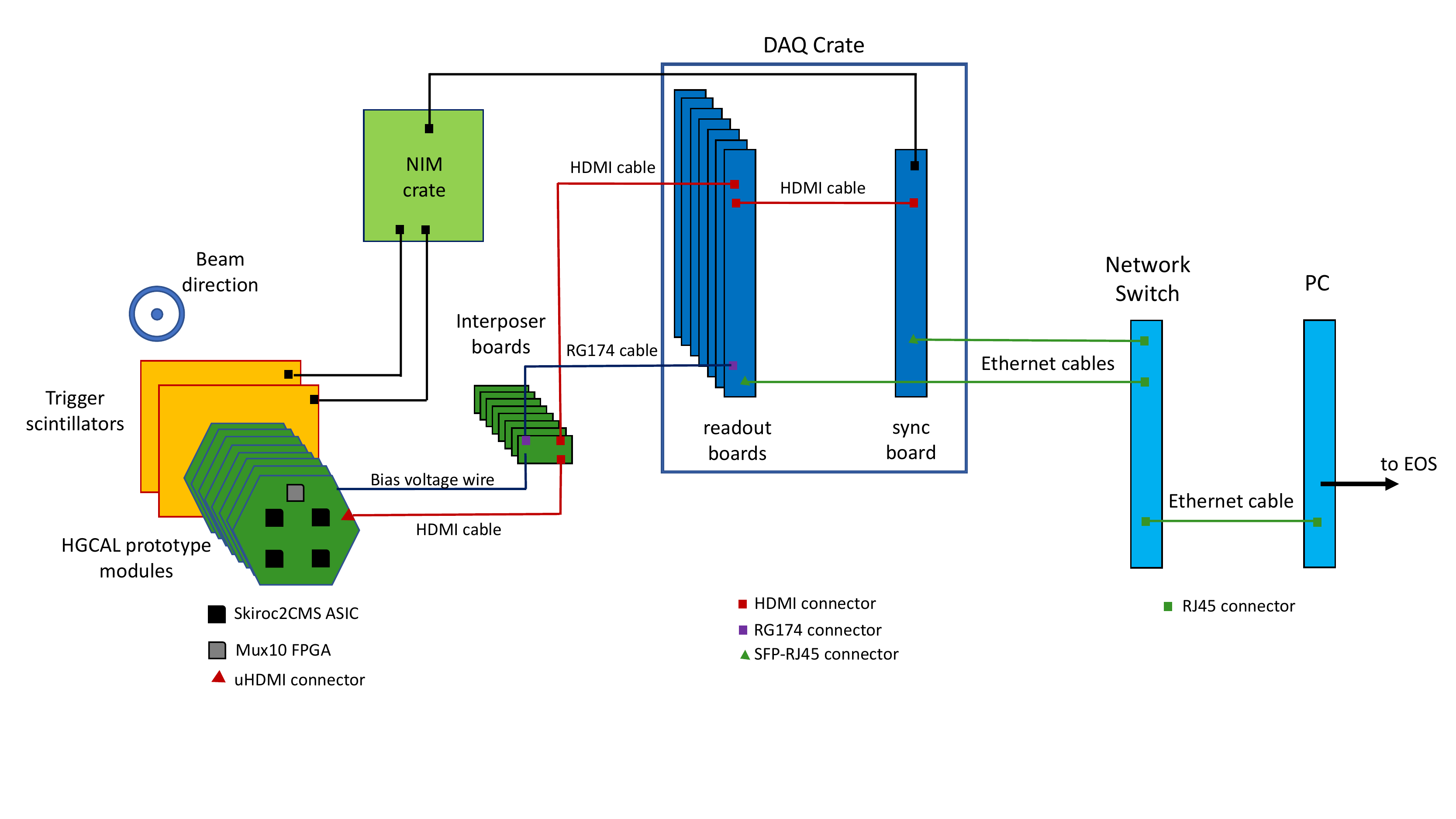}
\caption{\label{fig:DAQoverview} Schematic view of the DAQ system used in HGCAL beam tests.}
\end{figure}

Figure~\ref{fig:DAQoverview} shows a schematic view of the inter-connectivity of the DAQ system. The readout boards were connected to the prototype silicon modules by HDMI cables both between the readout boards and the interposer boards and between the interposer boards and the modules. The trigger signal was formed from a coincidence of signals from two scintillation counters located upstream of the calorimeter. The 40~MHz system clock, generated on the sync board and the trigger signals were transmitted to the readout boards with HDMI cables. Since the beam was not synchronised with the clock, the trigger and the clock were asynchronous. The bias voltage was distributed to the silicon sensors through the readout boards with separate RG174~cables.

\section{Data format and back-end DAQ components}
\label{sec:backend}

\subsection{Skiroc2-CMS ASIC and prototype module data format}
\label{subsec:skiroc2cms}

In operation, the analog signals were stored every 25~ns in a switched-capacitor array (SCA) with a depth of 13 cells. When a trigger was received, the updating of the SCA was halted and the two values for the ToA were stored, one referenced to the next falling edge of the 40 MHz clock, and the other to the next rising clock edge. Two values of the ToT were also kept, one with a fast ramp time-to-digital converter and another with a slow ramp.
\begin{figure}[htpb]
  \centering
  \begin{bytefield}[bitwidth=1.6em]{16}
    \bitbox[]{5}{} & \bitheader{0,1,2,3,4,15} \\
    \begin{rightwordgroup}{1924~$\times$~16-bit\\~~~~~integers}
      \bitbox[]{5}{$\times$64 channels} & \colorbitbox{lightblue}{1}{1} & \colorbitbox{lightblue}{1}{0} & \colorbitbox{lightblue}{1}{0} & \colorbitbox{lightblue}{1}{$H_{A}$} & \colorbitbox{lightblue}{12}{Low gain ADC (SCA0)}\\
      \bitbox[]{5}{$\times$64 channels} & \colorbitbox{lightcyan}{1}{1} & \colorbitbox{lightcyan}{1}{0} & \colorbitbox{lightcyan}{1}{0} & \colorbitbox{lightcyan}{1}{$H_{A}$} & \colorbitbox{lightcyan}{12}{High gain ADC (SCA0)}\\
      \bitbox[]{5}{} & \colorbitbox{gray!30!white}{16}{...$\times$13 SCA cells}\\
      \bitbox[]{5}{$\times$64 channels} & \colorbitbox{lightblue}{1}{1} & \colorbitbox{lightblue}{1}{0} & \colorbitbox{lightblue}{1}{0} & \colorbitbox{lightblue}{1}{$H_{A}$} & \colorbitbox{lightblue}{12}{Low gain ADC (SCA12)}\\
      \bitbox[]{5}{$\times$64 channels} & \colorbitbox{lightcyan}{1}{1} & \colorbitbox{lightcyan}{1}{0} & \colorbitbox{lightcyan}{1}{0} & \colorbitbox{lightcyan}{1}{$H_{A}$} & \colorbitbox{lightcyan}{12}{High gain ADC (SCA12)}\\
      \bitbox[]{5}{$\times$64 channels} & \colorbitbox{lightyellow}{1}{1} & \colorbitbox{lightyellow}{1}{0} & \colorbitbox{lightyellow}{1}{0} & \colorbitbox{lightyellow}{1}{$H_{A}$} & \colorbitbox{lightyellow}{12}{ToA (stop falling clk)}\\
      \bitbox[]{5}{$\times$64 channels} & \colorbitbox{lighteryellow}{1}{1} & \colorbitbox{lighteryellow}{1}{0} & \colorbitbox{lighteryellow}{1}{0} & \colorbitbox{lighteryellow}{1}{$H_{A}$} & \colorbitbox{lighteryellow}{12}{ToA (stop rising clk)}\\
      \bitbox[]{5}{$\times$64 channels} & \colorbitbox{lightmagenta}{1}{1} & \colorbitbox{lightmagenta}{1}{0} & \colorbitbox{lightmagenta}{1}{0} & \colorbitbox{lightmagenta}{1}{$H_{T}$} & \colorbitbox{lightmagenta}{12}{ToT (fast ramp)}\\
      \bitbox[]{5}{$\times$64 channels} & \colorbitbox{lightermagenta}{1}{1} & \colorbitbox{lightermagenta}{1}{0} & \colorbitbox{lightermagenta}{1}{0} & \colorbitbox{lightermagenta}{1}{$H_{T}$} & \colorbitbox{lightermagenta}{12}{ToT (slow ramp)}\\
      \bitbox[]{5}{} & \colorbitbox{lightorange}{1}{0} & \colorbitbox{lightorange}{1}{0} & \colorbitbox{lightorange}{1}{0} & \colorbitbox{lightorange}{13}{Roll position (13-bit)} \\
      \bitbox[]{5}{} & \colorbitbox{lightorange}{1}{0} & \colorbitbox{lightorange}{1}{0} & \colorbitbox{lightorange}{14}{Global timestamp MSB (14-bit)} \\
      \bitbox[]{5}{} & \colorbitbox{lightorange}{1}{0} & \colorbitbox{lightorange}{1}{0} & \colorbitbox{lightorange}{1}{0} & \colorbitbox{lightorange}{12}{Global timestamp LSB (12-bit)} & \colorbitbox{lightorange}{1}{0} \\
      \bitbox[]{5}{} & \colorbitbox{lightorange}{1}{1} & \colorbitbox{lightorange}{1}{1} & \colorbitbox{lightorange}{1}{0} & \colorbitbox{lightorange}{1}{0} & \colorbitbox{lightorange}{1}{0} & \colorbitbox{lightorange}{1}{0} & \colorbitbox{lightorange}{1}{0} & \colorbitbox{lightorange}{1}{0} & \colorbitbox{lightorange}{8}{Chip ID (8-bit)}
    \end{rightwordgroup}
  \end{bytefield}
  \caption{Data format of the Skiroc2-CMS ASIC. $H_{A}$ is the hit bit for ToA and is set to '1' when ToA is fired. Similarly, $H_{T}$ is the hit bit for ToT and is set to '1' when ToT is fired. The 13-bit roll position is used to reorder the SCA cells in time.}
  \label{fig:sk2cmsdata}
\end{figure}
Figure~\ref{fig:sk2cmsdata} shows the data format of the Skiroc2-CMS ASIC. 

When a trigger was received, by the MAX\textregistered10 FPGA of the hexaboard the four Skiroc2-CMS ASICs converted the data in analogue memory to the digital data format. These data were then read out by the MAX\textregistered10 FPGA and packaged as shown in Figure~\ref{fig:hbdata}. The bits $b_{Ai}$ belong to the ASIC "i" which followed the Skiroc2-CMS data format shown in Fig.~\ref{fig:sk2cmsdata}. For every event 30784 bytes of data were transmitted from each hexaboard. These data were then gathered, via the HDMI-uHDMI cables, by the back-end readout boards.
\begin{figure}[htbp]
    \centering
     \begin{bytefield}[bitwidth=3.0em]{8}
        \bitheader{0,3,4,7} \\
        \colorbitbox{lightcyan}{4}{HEADER} & \colorbitbox{lightcyan}{4}{1 bit per ASIC} \\
        \begin{rightwordgroup}{1924~$\times$~16 bytes}
        \bitboxes{1}{{1} {0} {0} {0} {$b_{A0}$} {$b_{A1}$} {$b_{A2}$} {$b_{A3}$}}\\
        \colorbitbox{gray!30!white}{8}{...}\\
       \bitboxes{1}{{1} {0} {0} {0} {$b_{A0}$} {$b_{A1}$} {$b_{A2}$} {$b_{A3}$}}
        \end{rightwordgroup}
    \end{bytefield}
    \caption{Hexaboard data format. The bits $b_{Ai}$ belong to the ASIC "i" which followed the Skiroc2-CMS data format described by Figure~\ref{fig:sk2cmsdata}. }
    \label{fig:hbdata}
\end{figure}

\subsection{Back-end DAQ electronics}


The DAQ system was designed to be easily scalable to provide a readout for different numbers of silicon modules. To minimise costs, readily available commercial components were used. Additionally, optical receiver modules (oRMs)~\cite{bib.Baber_2014}, recovered from the CMS level-1 trigger system, when it was upgraded with faster electronics, were used. Each oRM was equipped with a Kintex-7 FPGA, 4.8 Mbits of block RAM, two 6.6 Gbit/s bi-directional serial ports, and a 128 Mbit FLASH memory for configuration. The connection from the Kintex-7 FPGA to the gigabit Ethernet switch was made with SFP to RJ-45 adapters. 


\subsubsection{Readout board}
\label{subsec:rdoutboard}
In the final beam test in October 2018, there were 94 silicon modules that were read out with 14 readout boards mounted in two racks, controlled by a sync board. The readout boards performed the following tasks:
\begin{itemize}
    \item Loading firmware on the Max\textregistered10 FPGAs and module initialisation and reset.
    \item Generating and distributing control signals for the prototype silicon modules. 
    \item Accumulating the data received from the prototype silicon modules. 
    \item Distributing the clock, busy and trigger signals. 
    \item Distributing the low voltage power and the bias voltage for the prototype silicon modules.
\end{itemize}

\begin{figure}[htbp]
\centering
\includegraphics[width=1.1\textwidth,origin=c,angle=0]{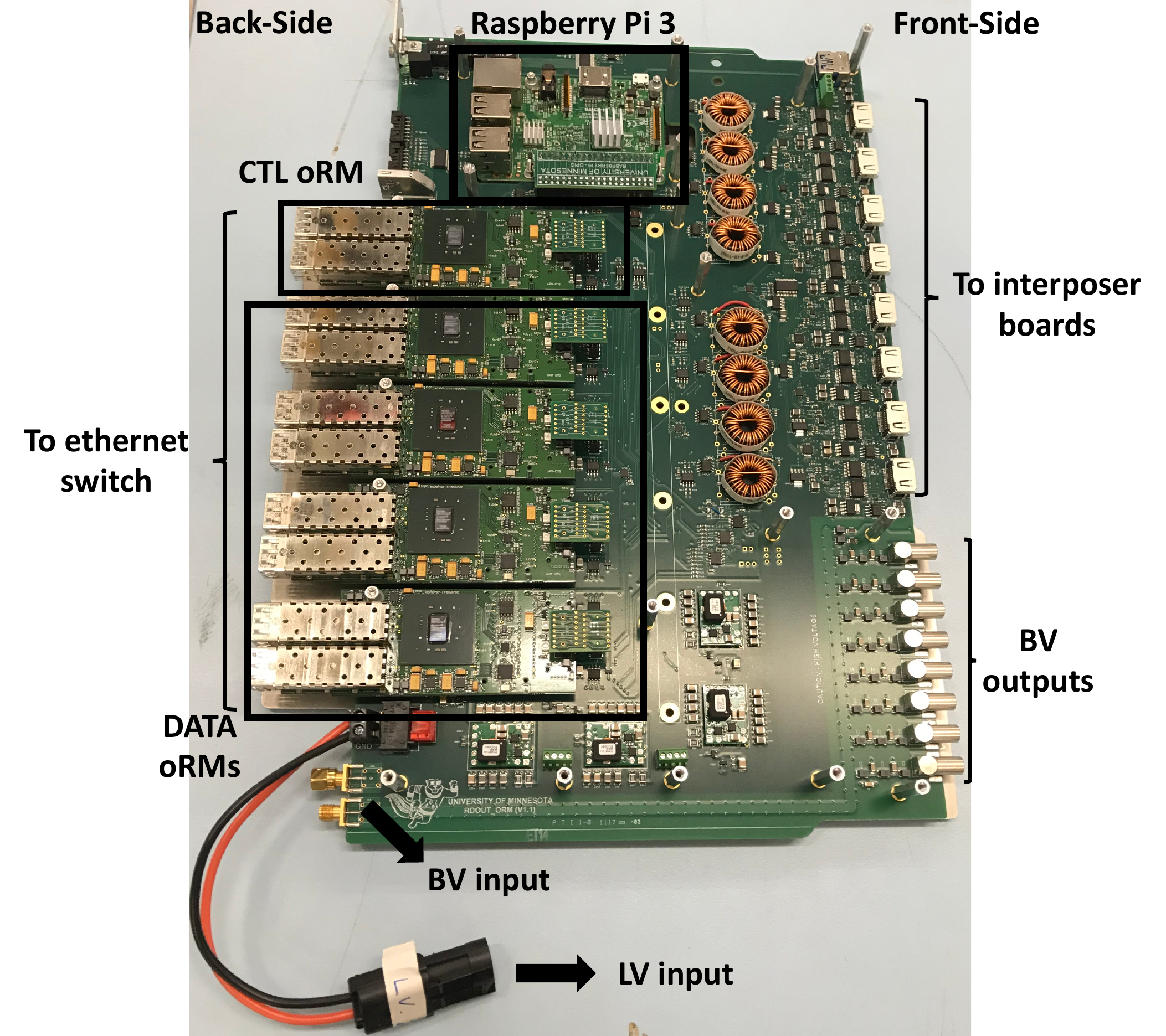}
\caption{\label{fig:i} The readout board used to send control data to the silicon modules, to receive data from them and to transmit it to the data acquisition computer. It readout data from up to eight silicon modules via HDMI connectors.  It also supplied the bias voltage for up to eight prototype silicon modules via standard RG174~connectors. It was equipped with one control and four data oRMs and one Raspberry Pi.}
\label{fig:Readout-Board}
\end{figure}

The readout board, shown in Figure~\ref{fig:Readout-Board}, was a custom PCB equipped with five oRMs and a Raspberry Pi. Each board had eight HDMI ports on the front panel connected to the silicon modules through the interposers and one HDMI port on the back panel for connection to the sync board. On the front of each board there were eight RG174~connectors that were used to distribute, after filtering, the bias voltage through the interposers to each detector module.


The overall readout cycle was controlled via helper processes running on the Raspberry Pi, which communicated with each oRM through the SPI bus. Each Pi was connected to the central DAQ server through its Ethernet port, from which it also received `Start', `Stop', and other commands.

A single readout board was equipped with five oRMs: one control (CTL) oRM, and four DATA oRMs. The DATA oRMs were responsible for reading the data from up to two silicon modules, while the CTL oRM received data from the DATA oRMs and transferred it to the central server. It also managed the communication with the sync board. 
The firmware installed on the CTL oRM's FPGA included the IPBus firmware~\cite{bib.Larrea_2015, bib.Mans_2010} for this purpose. The IPBus IP and MAC addresses of the oRM were set by the Raspberry Pi, as well as other parameters used by the CTL oRM as it combined the four streams of data.


Before a run started, the helper processes on the Raspberry Pis first configured the ASICs on the prototype silicon modules,
and data collection was initiated.
The Trigger signal was broadcast from the sync board to the readout boards, from where it was forwarded on to the modules. 
On the readout boards, the helper processes running on the Raspberry Pis after receiving the trigger signal, prompted the ASICs to initiate data transmission. The data from the ASICs were then sent unprocessed, via the MAX\textregistered10 FPGA, to the DATA oRMs, where it was then merged into a single data stream by the CTL oRM.
%
%
The 4-bit headers of Figure~\ref{fig:hbdata} were then dropped and 32-bit integers were built with the data from up to eight modules, corresponding to 32 ASICs. These 32-bit integers were written to a FIFO to be readout by the central server using the IPBus protocol over a gigabit Ethernet link.  When ready, a flag was set inside the CTL's RAM to indicate that the data were ready for transfer. 
In Fig.~\ref{fig:ctlormdataformat} the output data format of the CTL oRM FIFO is shown.

\begin{figure}[htbp]
\centering 
\includegraphics[width=1.0\textwidth,origin=c,angle=0]{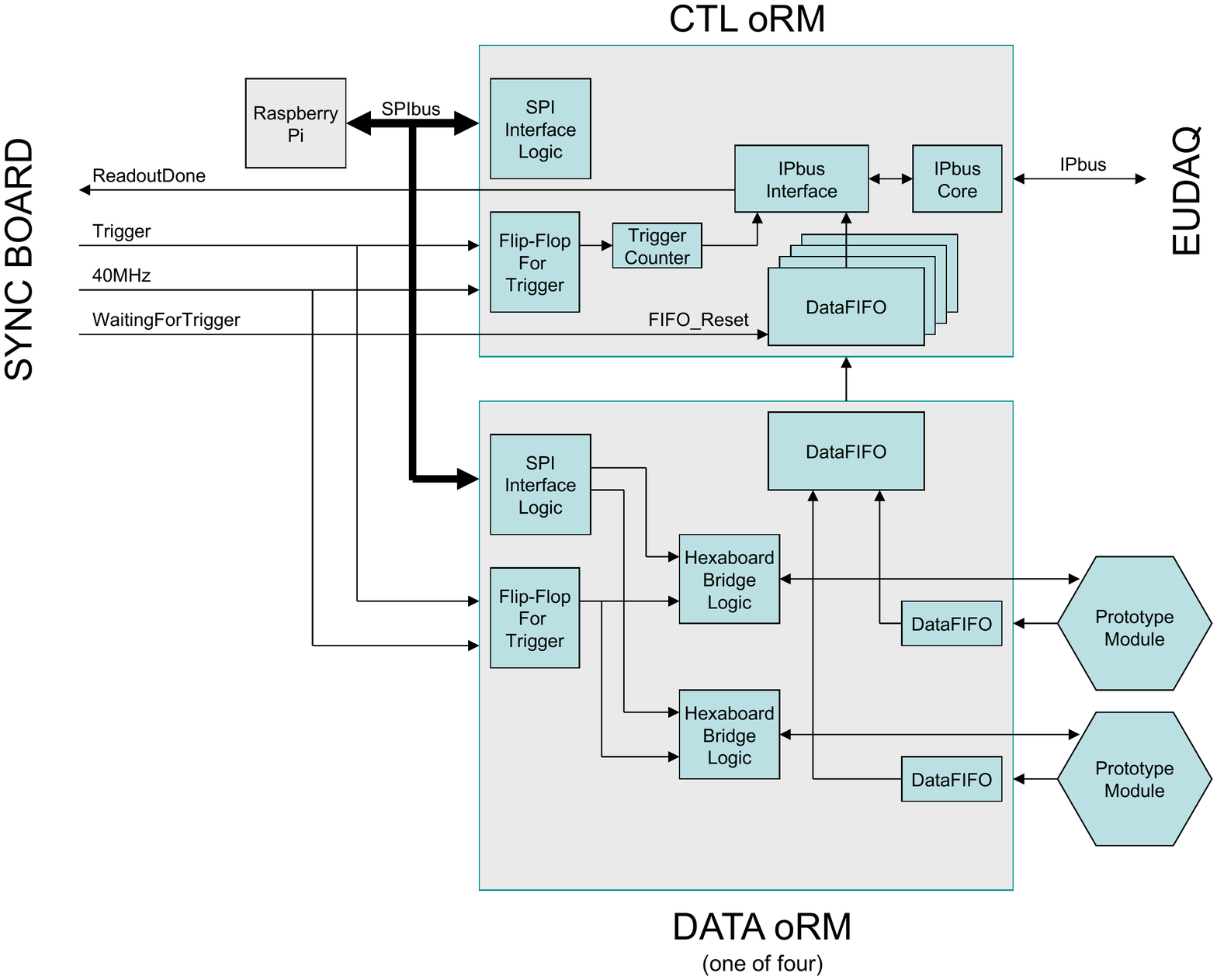}
\caption{\label{fig:i} The firmware block diagram of the readout board.}
\label{fig:ReadoutBoard_fw}
\end{figure}

Once the data had been fully read out by the server, the helper processes on the Pis reset the ASICs, and sent a start acquisition signal. The CTL oRM then sent a `ReadoutDone' signal to the sync board, indicating the boards had finished their cycles and were ready to receive the next trigger. The firmware block diagram of the readout board is shown in Figure~\ref{fig:ReadoutBoard_fw}.

\begin{figure}[htbp]
  \centering
  \begin{bytefield}[bitwidth=2.0em,bitformatting=\fakethirtytwobits]{16}
    \bitheader{0-15} \\
    \begin{rightwordgroup}{30784~$\times$~32-bit\\~~~~~integers}
      \bitboxes{1}{ {$b_{0,0}$} {$b_{0,1}$} {$b_{0,2}$} {$b_{0,3}$} }
      \colorbitbox{gray!30!white}{8}{...}
      \bitboxes{1}{{$b_{7,0}$} {$b_{7,1}$} {$b_{7,2}$} {$b_{7,3}$}
      }\\
      \colorbitbox{gray!30!white}{16}{...}\\
      \bitboxes{1}{ {$b_{0,0}$} {$b_{0,1}$} {$b_{0,2}$} {$b_{0,3}$} }
      \colorbitbox{gray!30!white}{8}{...}
      \bitboxes{1}{{$b_{7,0}$} {$b_{7,1}$} {$b_{7,2}$} {$b_{7,3}$}
      }
    \end{rightwordgroup}
  \end{bytefield}
  \caption{Data format of the CTL oRM FIFO readout by the central server using the IPBus protocol. The bits $b_{i,j}$ correspond to the data of ASIC "j" of module "i".}
  \label{fig:ctlormdataformat}
\end{figure}

\subsubsection{Sync board}

\begin{figure}[htbp]
\centering 
\includegraphics[width=1.1\textwidth,origin=c,angle=0]{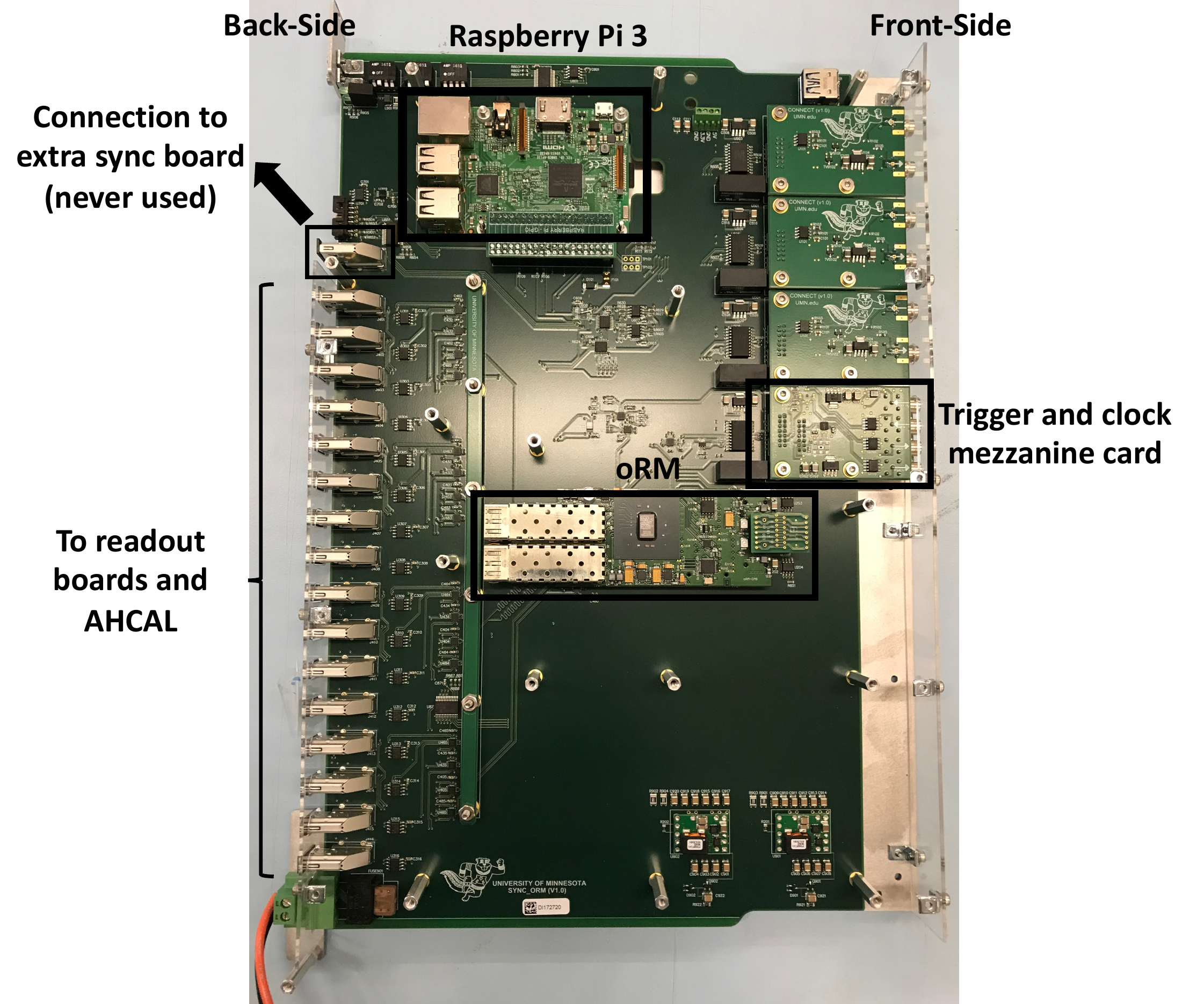}
\caption{\label{fig:i} The sync board was used to control up to 15 readout boards. It received the Veto and the Trigger signals and distributed control signals to up to 15 readout boards. On top of the 15 HDMI ports for readout board control there was one extra port for connection to another sync board for daisy-chaining. It was equipped with a Raspberry Pi and an Kintex-7 FPGA for control and communication. It was also equipped with a mezzanine card for clock generation and receiving external signals.}
\label{fig:Synch-Board}
\end{figure}

The function of the sync board, shown in Figure~\ref{fig:Synch-Board}, was the distribution of the common signals to the readout boards and to synchronise the flow of data from the readout boards.
The sync board generated the 40~MHz system clock on a small mezzanine card mounted at the rear of the board. 
On the same mezzanine there were four 50\,$\Omega$ RG174~connectors. Two were for the Trigger and a Veto signal inputs and two for the Clock and a copy of the Trigger signal outputs. The Veto signal was not used in these tests.
Processing on the sync boards was handled by a Raspberry Pi as well as a central (SYNC) oRM.
One sync board could control up to 15 readout boards through 15 HDMI ports mounted on the front. 
An extra HDMI port was mounted on the front to allow for the possibility of daisy-chaining two or more sync boards together when more than 15 readout boards are to be readout. 

\begin{figure}[htbp]
\centering 
\includegraphics[width=1.\textwidth,origin=c,angle=0]{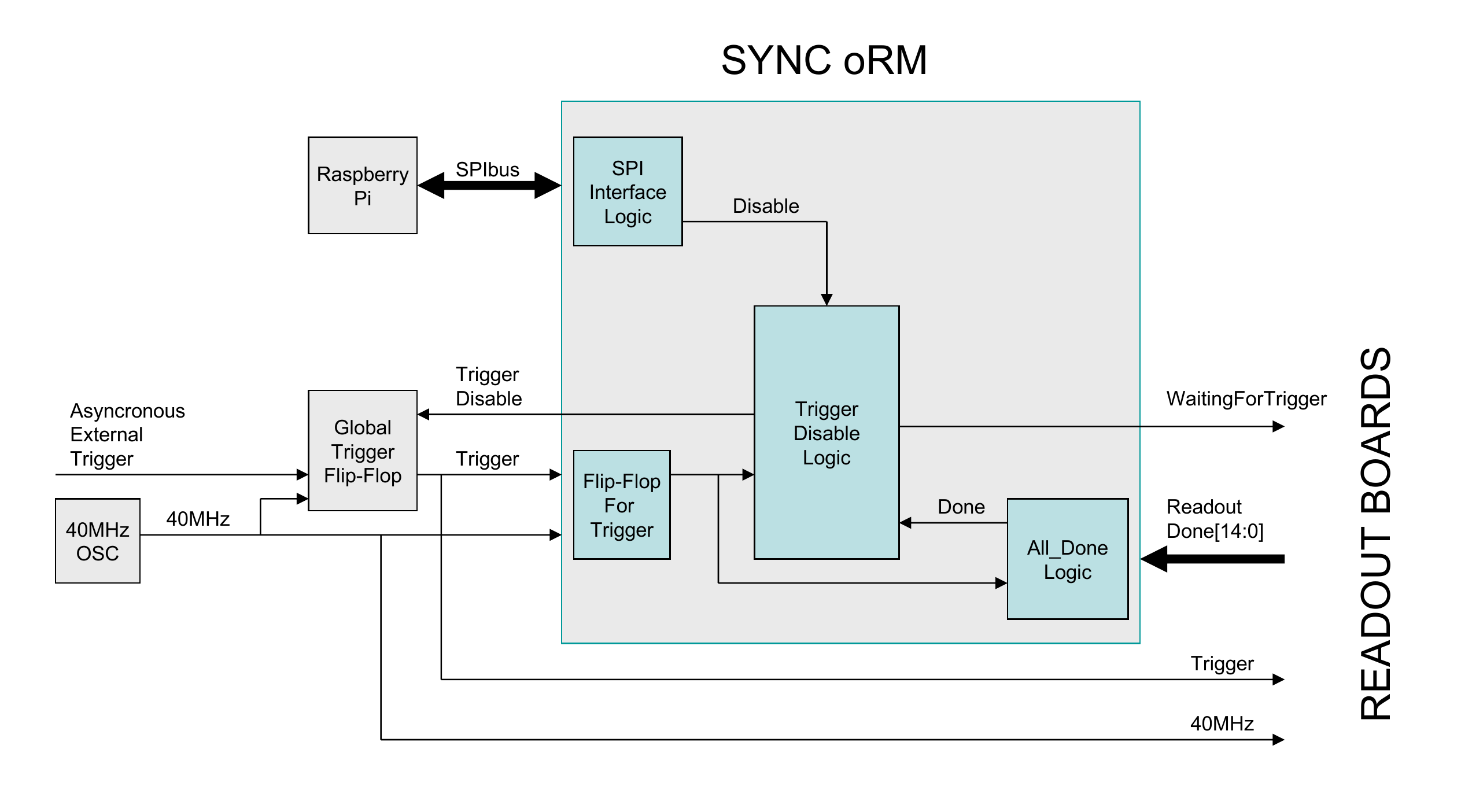}
\caption{\label{fig:i} The firmware block diagram of the sync board.}
\label{fig:SnycBoard_fw}
\end{figure}

At the start of a readout cycle, the sync board waited for an asynchronous External Trigger signal. Then this signal was synchronized with the on-board 40~MHz clock and sent to the readout boards to be distributed to the silicon modules. The sync board then waited for a `ReadoutDone' signal from each readout board. Once this signal was received from all the readout boards, the sync board made itself ready to process the next available trigger. The firmware block diagram of the sync board is shown in Figure~\ref{fig:SnycBoard_fw}.

\section{Data acquisition software}
\label{sec:software_framework}

The DAQ software selected for these tests was based on the EUDAQ~\cite{bib.eudaq} framework. This framework, written in C++, was developed specifically for small-to-medium scale systems with significantly less overhead than frameworks used in large scale experiments, like XDAQ~\cite{bib.xdaq}. Additionally, in separate earlier tests of the AHCAL prototype, the EUDAQ system had already been used successfully. 

The EUDAQ framework was designed to be modular and portable. It was structured so that software for the readout of specific detector components was kept separate and distinct from the core processes.  For this each detector component that produced data had a `Producer' process running. The functions performed by the Producer was to initialise, configure, issue stops and starts to the component, and to collect the data and forward it to the core process.   

\subsection{CMS-CE EUDAQ Producer}
\label{subsec:eudaq-ce-producer}

A Readout Producer was developed to read out the data from one or more readout boards in parallel. During the combined beam test of October 2018, the DAQ had seven of these Readout Producers, each connected to two readout boards. The $\mu$TCA Hardware Access Library ($\mu$HAL) was used to read the IPBus UDP transactions from the readout boards. The sequence of operations of the  Readout Producer were as follows: 

\begin{enumerate}
    \item Wait until each $\mu$HAL interface is notified that a trigger occurred (by checking an IPBus register of the CTL oRM boards).
    \item Read out the FIFO of the CTL oRM board from each readout board and fill raw data containers. The data format of this FIFO is described in Section~\ref{subsec:rdoutboard}. A time-stamp -- the number of 40 MHz clock cycles in a 64-bit integers since the last configuration -- is read out with the data.
    \item `ReadoutDone' signal is sent from CTL oRM to Sync oRM. 
    \item Create an event block containing the raw data from each readout board, the time-stamp of the readout board and the event ID. 
    \item Forward the event block to the EUDAQ data collector.
    \item Increment the event ID.
    \item Return to step 1 and wait for the next trigger.
\end{enumerate}

Once the readout was complete, data from each of the event blocks from each of the Readout Producers were combined with data from the Producers connected to other detector components to form a complete event data block. 

\subsubsection{EUDAQ online data monitoring}
Part of the EUDAQ framework were tools to monitor the data collection. Online analyses were developed to monitor in real time the stability of the pedestal values, noise and occupancies of each of the silicon sensor channels. 

\subsubsection{Data unpacking and first analysis steps}
The first step of the data analysis consisted of unpacking the CMS CE EUDAQ events. For this purpose a C++ library embedded in the EUDAQ framework has been developed. 

An initial data quality check was performed before unpacking the raw data by comparing the difference in time-stamp with the time-stamp of the previous event as a check of the synchronisation for all the readout boards.  During the October data taking only a few runs had events with a synchronization failure. 

After this test the data were unpacked and the data from each ASIC were sorted into tables of 16-bit integers with the structure shown in Figure~\ref{fig:sk2cmsdata} and stored in a ROOT~\cite{bib.root} file. As zero-suppression was not used for simplicity, 
these tables contained the data from every channel of the ASICs, including those not connected to a detector channel. 
The data for each cell in an event contained data and pointer information from the 13 SCA cells for both the high- and the low-gain slow shapers, the ToA and the ToT measurements. The pointer was the address of first SCA cell data for the event, which allowed the ordering of the SCA data in time. This was required for the data reconstruction since the trigger was asynchronous with the 40~MHz clock. The data analysis workflow, which was developed in the CMSSW framework~\cite{bib.cmssw}, and transformed the tables of 16-bit integers into a collection of calibarated hits for data analysis used the ROOT files as input. 
\section{System synchronization}
\label{sec:synchronization}
\subsection{Beam-characterization detectors}  
\label{sec:beammonitoring}

The tests of HGCAL and AHCAL prototypes with particle beams in the H2 area at CERN had been complemented by the readout of various beam-characterization detectors. Four delay wire chambers (DWC) \cite{bib.dwcs} measured the trajectory and impact of the particles in the beam, two scintillator detectors served as external trigger source and two micro-channel plates (MCP) had been used to provide fast signals for reference timing measurement of the incident particles \cite{bib.hgcal-beamtest-paper}. For this purpose, two 16-channel CAEN v1290N TDCs and one v1742 digitiser were integrated into the HGCAL prototype DAQ.

From each DWC four signals were separately discriminated at a threshold of -30~mV and fed as inputs to the TDC. Since the binning of the time-stamp digitization should be less than 1~ns corresponding to an optimal resolution of the position measurement of 200$~\mu$m  \cite{bib.dwcs}, a binning of 25~ps had been chosen. For proper event synchronization, the trigger for all CAEN modules stemmed from the duplicated TTL trigger signal issued by the synchronization board. After conversion to NIM, it was copied three times and fed into each module individually. After receiving a trigger, events were built, were labeled with trigger time-stamps and subsequently stored in a local buffer.

Two dedicated EUDAQ producers \cite{bib.eudaq} had been developed. They ran on a separate computer and communicated to these modules through optical link and VMEbus. These producers polled the event data from the buffers at a (configurable) frequency of 500$~\text{Hz}$ (HGCAL DAQ rate $\approx 40~\text{Hz}$), converted the raw data into the EUDAQ format and sent it to the main DAQ for storage and online data monitoring. 

\subsection{AHCAL} 

\begin{figure}[htbp]
  \centering
  \includegraphics[width=0.8\linewidth]{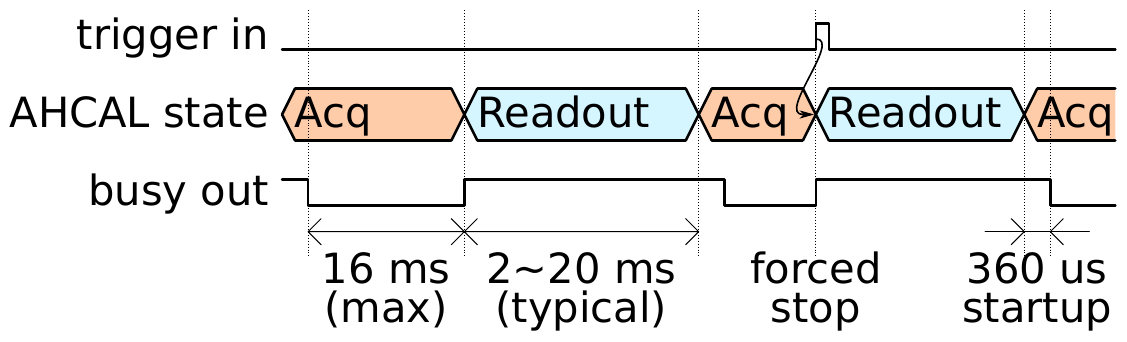}
  \caption{\label{fig:AHCAL-timing} The time diagram of AHCAL acquisition and readout phases.} 
\end{figure}

The front-end electronics of the AHCAL prototype were designed for low power operation under a specific ILC accelerator timing with a less than 1~ms long spill followed by 199~ms idle time. The operation of the Spiroc ASIC~\cite{bib.spiroc} was therefore split in 3 phases: 1) acquisition phase, where self-triggered events were stored into up to 16 analog memory columns; 2) conversion phase, where up to 16 events were sequentially digitized by internal ADC; 3) readout phase, where the digitized data was read out. Detailed timing was described in~\cite{bib.ahcal_daq}. For beam test purposes, the acquisition phase length was extended to 16~ms and the readout phase varied typically between 2 to 20~ms, depending on the hit occupancy. Any external trigger from the sync board stopped the acquisition for immediate readout, as shown in Fig~\ref{fig:AHCAL-timing}.

Due to the `self-trigger' design of the ASIC, the AHCAL did not require an external trigger for data taking. All the hits (including noise hits) were read out and referenced by a number of bunch-crossing clock cycles (4~us period, called BXID) from the start of the acquisition phase. In order to assign an external trigger to the hits in the AHCAL, the DAQ internally samples the external trigger number (with a time-stamp, 48-bit counter with a 25~ns resolution) and the time-stamp of the start of the relevant acquisition phase. The trigger was assigned to one of the self-triggered events in the acquisition cycle. Assignment was based on the startup time from the start of the acquisition phase to the beginning of the first BXID and the additional delay due to the length of the trigger cable. 

\begin{figure}[htbp]
  \centering
  \includegraphics[width=0.8\linewidth]{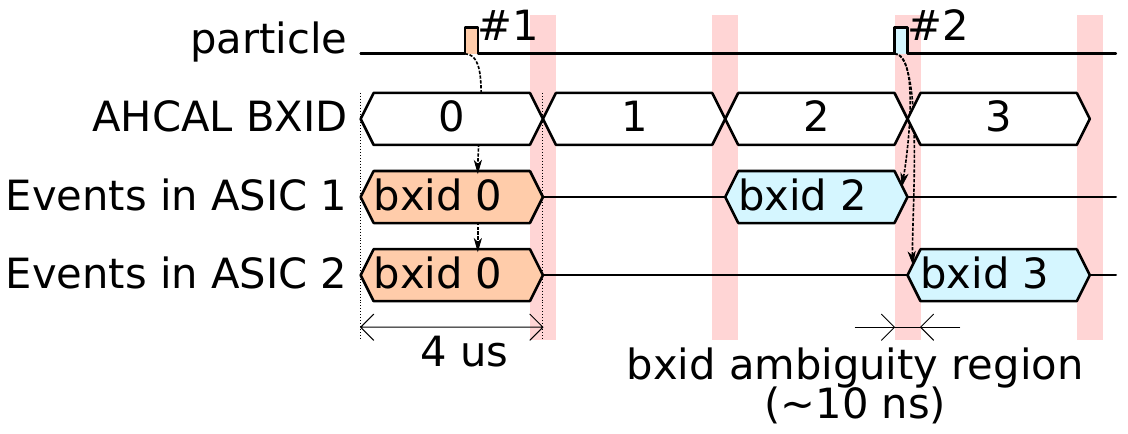}
  \caption{\label{fig:AHCAL-bxid} The organization of self-triggered particles in AHCAL BXIDs, showing the BXID ambiguity for an example of split events due to the time of arrival with respect to the BXID clock.} 
\end{figure}

The collation of events according to the BXID might have however led to an existence of incomplete events for particles, that arrived close to the BXID counter value switching in the ASICs. An example for such a split event is shown in Figure~\ref{fig:AHCAL-bxid} (particle no. 2). Several factors contribute to the BXID ambiguity: time walk of the signals, clock skew due to board and ASIC location, clock tree distribution through the FPGAs and clock jitter. The internal ASIC TDC~\cite{bib.spiroc} had also a region of non-linearity around the BXID change. Therefore, particles arriving close to the BXID change, within 10~ns,  needed to be excluded from the data analysis.

The AHCAL provided two means of synchronization with the HGCAL data: the trigger number and the trigger time-stamp, which used the 40~MHz clock from the HGCAL sync board. Both pieces of information were accessible in the data file. 
\section{Collected Data}
\label{sec:performance}

The DAQ was operated at a readout rate of 40~Hz. At this rate the amount of data readout for a typical 5 second spill of the SPS for the HGCAL prototype was approx 300 MB, as each of the 94 silicon modules sent 16 KB of data per event.

\begin{figure}[htbp]
  \centering
  \includegraphics[width=.85\linewidth]{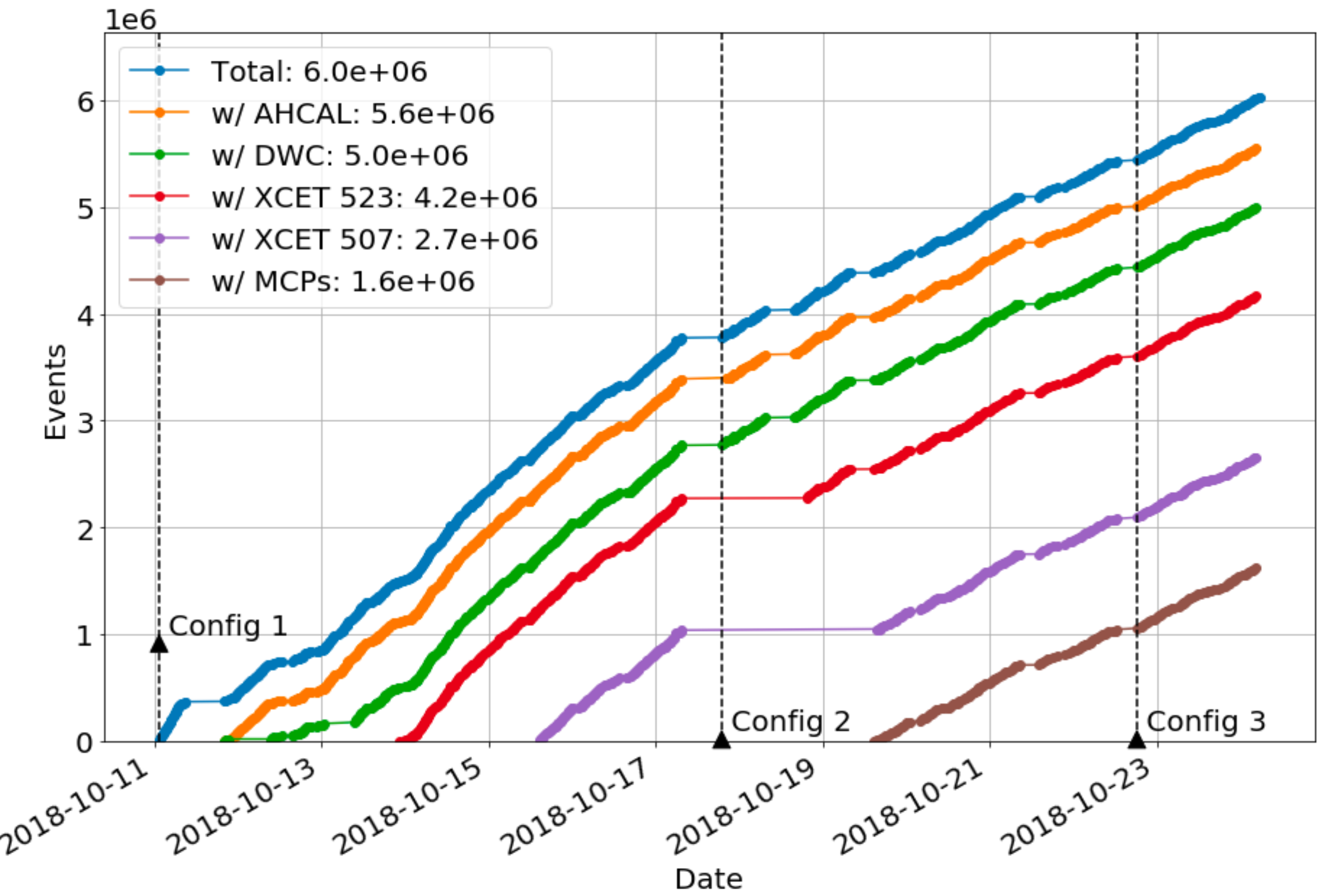}
  \caption{\label{fig:performance} The accumulated events for different detectors during beam-test campaign in October 2018.}
\end{figure}

In the last run, over a period of two weeks, six million events were collected with beams of charged hadrons, electrons, and muons with momenta from 20 to 300 GeV/c, with different detector configurations. Figure~\ref{fig:performance} shows accumulated events for different detectors over the beam-test campaign. 

\section{Summary}
\label{sec:conclusion}

In the upgrade of the CMS detector for when the HL-LHC is operational, the two endcap calorimeters will be replaced with high granularity sampling calorimeters equipped with silicon sensors. As part of the development of this calorimeter, a series of beam tests have been conducted with different sampling configurations using prototype segmented silicon detectors readout with a low-cost custom scalable data acquisition system. 
%
%
The software framework used for the run control and data collection was the portable modular EUDAQ framework. 
In the most recent of the tests conducted in late 2018 at the CERN SPS in 2018, the performance of a prototype calorimeter equipped with ${\approx}12,000\rm{~channels}$ of silicon sensors, in conjunction with the CALICE prototype analogue hadron calorimeter, was studied with beams of high-energy electrons, pions and muons, with six million events collected over a two week period.
\acknowledgments
We thank the technical and administrative staffs at CERN and at other CMS institutes for their contributions to the success of the CMS effort.  We acknowledge the enduring support provided by the following funding agencies: BMBWF and FWF (Austria); CERN; CAS, MoST, and NSFC (China); MSES and CSF (Croatia); CEA and CNRS/IN2P3 (France); SRNSF (Georgia);  BMBF, DFG, and HGF (Germany); GSRT (Greece); DAE and DST (India); MES (Latvia); MOE and UM (Malaysia); MOS (Montenegro); PAEC (Pakistan); FCT (Portugal); JINR (Dubna); MON, RosAtom, RAS, RFBR, and NRC KI (Russia); MST (Taipei); ThEPCenter, IPST, STAR, and NSTDA (Thailand); TUBITAK and TENMAK (Turkey); STFC (United Kingdom); DOE (USA).



\bibliography{ref}
\bibliographystyle{ieeetr}

\end{document}